 \documentclass[final,3p,times,twocolumn]{elsarticle}

\usepackage[T1]{fontenc}
\usepackage{bm}
\usepackage{color}
\usepackage{graphicx}
\usepackage{hyperref}
\usepackage{amsmath,empheq}
\usepackage{amssymb}
\usepackage{amsfonts}
\usepackage{overpic}
\usepackage[table,xcdraw]{xcolor}

\newcounter{bla}

\journal{Computer Physics Communications}

\newcommand{\unit}[1]{\hat{\boldsymbol{\mathrm{#1}}}}

\newcommand{\thb}{\Theta_b}
\newcommand{\besselK}{\mathrm{K}}
\newcommand{\W}{\bm{W}}
\newcommand{\Y}{\bm{Y}}
\newcommand{\y}{Y}
\newcommand{\tm}{t_-}
\newcommand{\tp}{t_+}
\newcommand{\eps}{\varepsilon_0}

\newcommand{\etol}{\varepsilon_\mathrm{tol}}
\newcommand{\eabs}[1]{\varepsilon_\mathrm{abs,#1}}

\newcommand{\edrift}{\varepsilon_\mathrm{drift}}
\newcommand{\ediff}[1]{\varepsilon_{\mathrm{diff},#1}}

\newcommand{\gD}{\mathcal{D}}
\newcommand{\gJ}{\mathcal{J}}

\newcommand{\gZv}{\mathbf{Z}}

\newcommand{\FEP}{FEP}
\newcommand{\AMP}{AMP}
\newcommand{\FMG}{FMG}
\newcommand{\AMG}{AMG}

\DeclareMathOperator{\sign}{sgn}

\begin{document}

\begin{frontmatter}



\title{Adaptive time-stepping Monte Carlo integration of Coulomb collisions}


\author[1]{K. S\"arkim\"aki\corref{author}}
\author[2]{E. Hirvijoki}
\author[1]{J. Ter\"av\"a}

\cortext[author] {Corresponding author.\\\textit{E-mail address:} konsta.sarkimaki@aalto.fi}
\address[1]{Aalto University, Espoo, Finland}
\address[2]{Princeton Plasma Physics Laboratory, Princeton, NJ, USA}

\begin{abstract}
We report an accessible and robust tool for evaluating the effects of Coulomb collisions on a test particle in a plasma that obeys Maxwell-J\"uttner statistics. The implementation is based on the Beliaev-Budker collision integral which allows both the test particle and the background plasma to be relativistic. The integration method supports adaptive time stepping, which is shown to greatly improve the computational efficiency. The Monte Carlo method is implemented for both the three-dimensional particle momentum space and the five-dimensional guiding center phase space.

Detailed description is provided for both the physics and implementation of the operator. The focus is in adaptive integration of stochastic differential equations, which is an overlooked aspect among existing Monte Carlo implementations of Coulomb collision operators. We verify that our operator converges to known analytical results and demonstrate that careless implementation of the adaptive time step can lead to severely erroneous results.

The operator is provided as a self-contained Fortran 95 module and can be included into existing orbit-following tools that trace either the full Larmor motion or the guiding center dynamics. The adaptive time-stepping algorithm is expected to be useful in situations where the collision frequencies vary greatly over the course of a simulation. Examples include the slowing-down of fusion products or other fast ions, and the Dreicer generation of runaway electrons as well as the generation of fast ions or electrons with ion or electron cyclotron resonance heating.





\end{abstract}

\begin{keyword}
Coulomb collision\sep Monte Carlo\sep Fokker-Planck equation\sep Milstein method

\end{keyword}

\end{frontmatter}



{\bf PROGRAM SUMMARY}

\begin{small}
\noindent
{\em Program Title:} AMCC -- (A)daptive (M)onte-(C)arlo (C)oulomb collisions\\
{\em Licensing provisions:} LGPL-2.1                              \\
{\em Programming language:} Fortran 95                          \\
{\em Nature of problem:}
Test-particle tracing is a common feat within existing fusion applications. 
While efficient adaptive methods exist for integrating the incompressible Hamiltonian flow, the effects of Coulomb collisions are commonly implemented with far less sophisticated algorithms. 
\\
{\em Solution method:}
The relativistic Fokker-Planck equation for test-particles in Maxwell-J\"uttner background plasmas is converted into a stochastic differential equation. 
The stochastic differential equation is solved using adaptive Monte Carlo techniques. 
Methods to evaluate the effect of Coulomb collisions for both the three-dimensional particle momentum space and the five-dimensional reduced guiding center phase space are included. \\
{\em Additional comments including Restrictions and Unusual features:}\\
The package includes optionality to evaluate the relativistic Fokker-Planck coefficients, a feature useful for constructing accurate orbit averaged collision operators. 
The package also provides explicit one-step sympletic integrator for the relativistic Lorentz force that can be used for tracing test-particles in given electromagnetic backgrounds.
   \\

\end{small}

\section{Introduction}
\label{sec:intro}
Standard Runge-Kutta methods developed for ordinary differential equations (ODEs), are based on first converting the ODE into an integral equation, and then discretizing the integral. 
Stochastic differential equations (SDEs), often encountered when stochastic processes such as collisions are of interest, can equivalently be transformed into integral equations, with the exception that the resulting integral is non-Riemannian~\cite{oksendal2013stochastic}.
The discretization rules used for ODEs then no longer apply and, instead, either It\^o or Stratonovich calculus must be adopted to obtain numerical methods for solving SDEs.

Perhaps the most straight-forward discretization method for SDEs is the Euler-Maruyama method. 
While simple to implement, the Euler-Maruyama method suffers from constant time-step requirement in the sense that one is not allowed to simply recompute the step if the estimated integration error turns out too large, as this might lead to a convergence to a wrong solution~\cite{gaines1997variable}.
The reason is that while Euler-Maruyama has weak order of convergence 1.0, its strong order of convergence is only 0.5.
For adaptive integration of SDEs, a method must have strong convergence of at least 1.0 to guarantee that it converges to a correct solution.
A common application, where time steps are tested and rejected when needed, is test-particle tracing where the Hamiltonian motion of the particle is solved with adaptive Runge-Kutta methods.
Therefore, to treat collisions with a manner consistent with the Hamiltonian motion, a higher order scheme than the Euler-Maruyama must be used.

Further, the test-particle collision frequencies in a plasma can vary greatly during the course of following the particle trajectory: energetic particles start as nearly collisionless and slowly drift toward the more collisional bulk population, while some particles from the bulk may accelerate to high energies. 
Therefore, adaptive time-stepping methods should be adopted also for the stochastic contribution to the particle motion, to reduce both the convergence errors and computational costs.

We thus see fit to introduce an adaptive time-stepping algorithm that would allow incorporating the Coulomb scattering into tracing of either the particle or guiding-center dynamics. 
Our Monte Carlo implementation evaluates the momentum change of a test-particle when it collides with a Maxwell-J\"uttner background plasma, consisting of electrons and possibly multiple ion species. 
The operator is based on the Beliaev-Budker collision integral~\cite{braams1987differential} and, therefore, is applicable even if either or both the test particle and the background plasma populations are relativistic.
Advanced Monte Carlo Coulomb collision operators have been developed quite recently~\cite{rosin2014multilevel,dimits2013higher}, but the operator developed here differs from those in that it is relativistic, adaptive, and also applicable in guiding center dynamics.

The paper is organized as follows.
In section~\ref{sec:test particle equation}, we begin by discussing how the Monte Carlo operator is obtained from the Beliaev-Budker collision integral by finding the related SDE.
The corresponding guiding-center operator is presented in section~\ref{sec:gc}.
Discretization of the relativistic particle collision operator is first done with the Euler-Maruyama method, in section~\ref{sec:MC operator}, where we also show how to compute the collision coefficients efficiently as these are also needed for the adaptive time-stepping. 
The adaptive scheme is described in detail in section~\ref{sec:adaptive} for both particle and guiding center picture. 
The collision operators are verified in section~\ref{sec:verification}, where we also compare the adaptive method with the fixed time step scheme, and confirm that the adaptive method is both faster and more accurate than the commonly used Euler-Maruyama method.

\section{Test-particle Fokker-Planck equation}
\label{sec:test particle equation}
In the limit of binary collisions, the collisional evolution of the particle distribution function of species $a$, interacting with species $b$, is determined by 
\begin{equation}
\label{eq:collision operator C_a}
\frac{\partial f_a}{\partial t}=\sum_bC_{ab}[f_a,f_b].
\end{equation}
In a plasma, the collisions are dominated by the small angle scattering events, so that the collision operator can be written in the Landau-Fokker-Planck form~\cite{Landau1937}
\begin{equation}
\label{eq:beliaev_budker_integral}
C_{ab}[f_a,f_b]=\frac{\Gamma_{ab}}{2m_{a}}\frac{\partial}{\partial\bm{u}}\cdot\int_{\mathbb{R}^3} d\bm{\bar{u}}\;\mathbf{U}_{\mathrm{BB}}\cdot\left(\frac{\bar{f}_{b}}{m_a}\frac{\partial f_a}{\partial\bm{u}} -\frac{f_a}{m_{b}}\frac{\partial \bar{f}_{b}}{\partial\bm{\bar{u}}}\right),
\end{equation}
where $\mathbf{U}_{\mathrm{BB}}$ is the relativistic Beliaev-Budker tensor.
The coordinate $\bm{u}=\bm{p}/m_ac$ is particle $a$ momentum normalized to the rest mass $m_a$ and speed of light $c$, while $\bm{\bar{u}}=\bm{\bar{p}}/m_bc$ denotes the same for particle of species $b$. The quantities with an overbar are evaluated at $\bm{\bar{u}}$. 
The operator also depends on the species charges, $q_a$ and $q_b$, through the coefficient $\Gamma_{ab}=q^2_aq^2_b\ln\Lambda_{ab}/(4\pi\varepsilon_0^2)$ where $\varepsilon_0$ is the vacuum permittivity, and $\ln\Lambda$ is the Coulomb logarithm. 
The Coulomb logarithm describes the ratio of minimum and maximum impact parameters $\ln\Lambda = \ln r_\mathrm{max}/r_\mathrm{min}$ and as such it indicates by which factor small angle scattering dominates the large angle scattering.
Natural choice for the maximum impact parameter is the Debye length , $\lambda_D = \sqrt{\eps \sum_b\left(T_b/n_bq_b^2\right)}$, where $n$ is density and $T$ is temperature. 
Depending which one is smaller, the minimum impact parameter is determined either via classical electron radius, $r_\mathrm{cl} = q_a q_b/(4\pi\eps m_r \tilde{v}^2)$, where $m_r=m_a m_b/(m_a+m_b)$ is the reduced mass and $\tilde{v} = \left<|\bm{v}_a-\bm{v}_b|\right>$ is the mean relative velocity, or by a quantum mechanical limit, $r_\mathrm{qm} = \hbar/(2m_r \tilde{v})$, where $\hbar$ is the reduced Planck constant.
In fusion plasmas, $\ln\Lambda$ has values in the range of 10 -- 20, but the uncertainties in estimating $r_\mathrm{min}$ means that the Coulomb logarithm is only accurate to within $1/\ln\Lambda$.

Returning to the Beliaev-Budker tensor, we find that it has a rather complicated expression~\cite{braams1987differential} 
\begin{equation}
\label{eq:beliaev_budker_tensor}
\mathbf{U}_{\mathrm{BB}}=\frac{r^2}{\bar{\gamma}\gamma w^3}\left(w^2\mathbf{I}-\bm{u}\bm{u}-\bm{\bar{u}}\bm{\bar{u}}+r(\bm{u}\bm{\bar{u}}+\bm{\bar{u}}\bm{u})\right),
\end{equation}
with $r=\gamma\bar{\gamma}-\bm{u}\cdot\bm{\bar{u}}$, $w=\sqrt{r^2-1}$, and $\gamma=\sqrt{1+u^2}$ is the Lorentz factor. 
In the non-relativistic limit, $c\rightarrow \infty$, the Beliaev-Budker tensor reduces to the better known Landau tensor~\cite{hinton1983collisional} 
\begin{equation}
\label{eq:landau_tensor}
\mathbf{U}_{\mathrm{L}}=\frac{1}{\lvert\bm{u}-\bm{\bar{u}}\rvert}\left(\mathbf{I}-\frac{(\bm{u}-\bm{\bar{u}})(\bm{u}-\bm{\bar{u}})}{\lvert\bm{u}-\bm{\bar{u}}\rvert^2}\right).
\end{equation}
From now on, we will focus solely on the relativistic expressions which, of course, are valid also in the non-relativistic regime.
In order to obtain the test particle collision operator, we first write the Beliaev-Budker collision integral, Eq.~\eqref{eq:beliaev_budker_integral}, in an explicit Fokker-Planck form
\begin{equation}
\label{eq:fokker-planck}
\frac{\partial f_a}{\partial t}=-\frac{\partial}{\partial\bm{u}}\cdot\left(\bm{K}_af_a\right)+\frac{\partial}{\partial\bm{u}}\frac{\partial}{\partial\bm{u}}:\left(\mathbf{D}_af_a\right),
\end{equation}
where the vector $\bm{K}_a=\sum_b\bm{K}_{ab}[f_b]$ and the tensor $\mathbf{D}_a=\sum_b\mathbf{D}_{ab}[f_b]$ are summations of the species-wise coefficients $\bm{K}_{ab}[f_b]$ and $\mathbf{D}_{ab}[f_b]$, and functionals of the distribution $f_b$. The expressions for the species-wise coefficients are
\begin{align}
\mathbf{D}_{ab}[f_b]&=\frac{\Gamma_{ab}}{2m_a^2}\int_{\mathbb{R}^3} d\bm{\bar{u}}\;\mathbf{U}_{\mathrm{BB}} \bar{f}_b,\\
\bm{K}_{ab}[f_b]&=\frac{m_a}{m_b}\frac{\Gamma_{ab}}{2m_a^2}\int_{\mathbb{R}^3} d\bm{\bar{u}}\;\mathbf{U}_{\mathrm{BB}}\cdot\frac{\partial\bar{f}_b}{\partial \bm{\bar{u}}}+ \frac{\partial}{\partial\bm{u}}\cdot\mathbf{D}_{ab}[f_b].
\end{align}

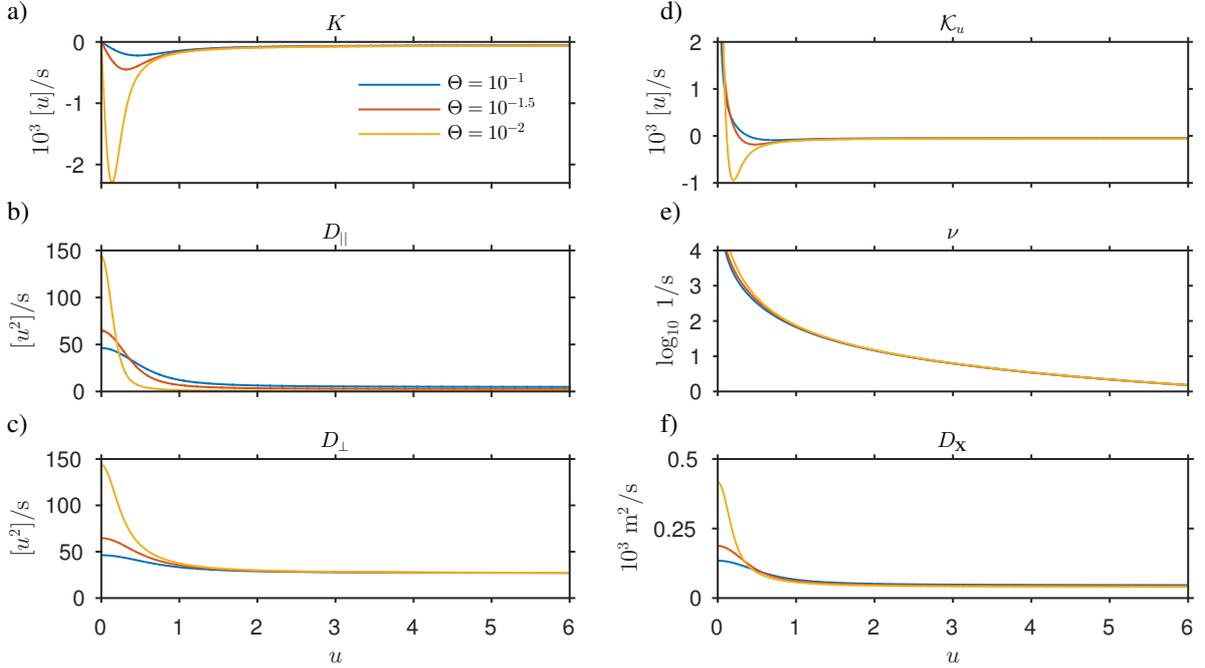
\begin{figure*}[!t]
\centering
\begin{overpic}[width=0.95\textwidth]{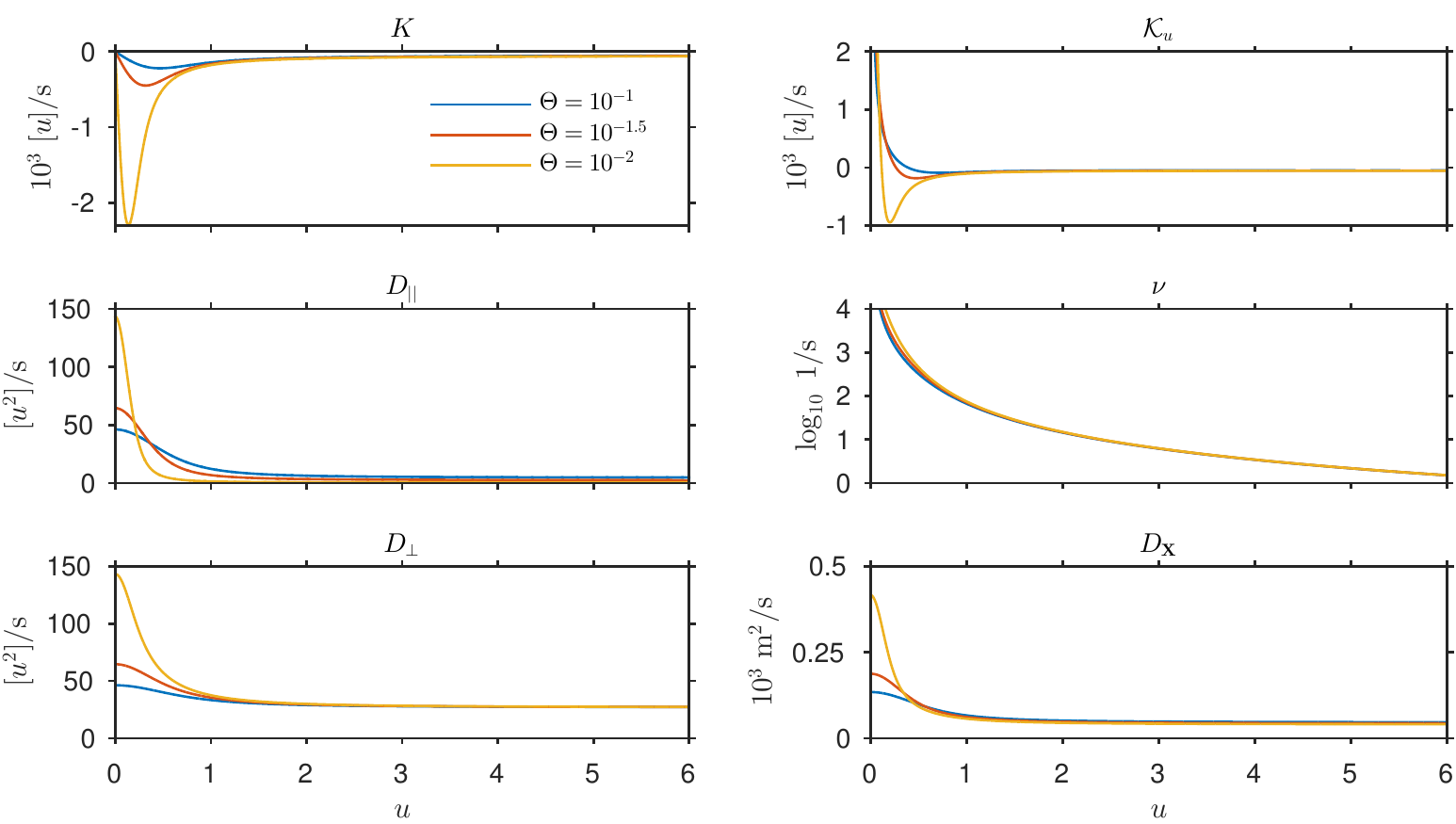}
\put(0,55){a)}
\put(0,38){b)}
\put(0,20){c)}
\put(55,55){d)}
\put(55,38){e)}
\put(55,20){f)}
\end{overpic}
\caption{Collision coefficients as a function of normalized momentum $u=p/mc$ for different values of normalized temperature $\Theta=T/mc^2$. 
(a)~--~(c) The coefficients related to the particle collision operator, Eq.~\eqref{eq:langevin equation for u_vec}.
(d)~--~(f) The coefficient related to the guiding center collision operator Eqs.~\eqref{eq: SDE GC X}~--~\eqref{eq: SDE GC u}. 
The values are for an electron test particle interacting with an electron plasma with density $n =10^{20}$ m$^{-3}$.
Note the logarithmic scale in (e).
}
\label{fig:coefficients}
\end{figure*}

From now on, we will assume that the distributions $f_b$ are Maxwell-J\"uttner distributions
\begin{equation}
\label{eq:maxwell juttner}
f_b(\bm{u})=\frac{n_b e^{-\sqrt{1+u^2}/\thb}}{4\pi \thb \besselK_2(1/\thb)}.
\end{equation}
Here $\besselK_{\nu}(x)$ (not to be confused with the coefficient $\bm{K}_{ab}$) is the $\nu^\mathrm{th}$ order modified Bessel function of the second kind, and $\Theta_b=T_b/m_bc^2$ is the normalized temperature.
In this case, the species-wise diffusion tensor $\mathbf{D}_{ab}$ and the force $\bm{K}_{ab}$ become isotropic~\cite{hinton1983collisional}
\begin{align}
\mathbf{D}_{ab}&=D_{ab,\parallel}(u)\bm{\hat{u}}\bm{\hat{u}}+D_{ab,\perp}(u)\left(\mathbf{I}-\bm{\hat{u}}\bm{\hat{u}}\right),\\
\bm{K}_{ab}&=K_{ab}(u)\bm{\hat{u}},
\end{align}
where $\bm{\hat{u}}\equiv\bm{u}/u$ is the unit vector parallel to $\bm{u}$. 
The coefficients $K_{ab}$, $D_{ab,\parallel}$, and $D_{ab,\perp}$ are defined in terms of three special functions $\mu_0(u;\Theta_b)$, $\mu_1(u;\Theta_b)$, and $\mu_2(u;\Theta_b)$, and they are given by the following expressions~\cite{Pike_Rose_2014}
\begin{align}
\label{eq:K}
K_{ab}&=-\frac{\Gamma_{ab}n_b}{m_a^2c^3}\frac{1}{u^2}\left(\frac{\mu_{0}}{\gamma}+\frac{m_a}{m_b}\mu_{1}\right),\\
\label{eq:Dpar}
D_{ab,\parallel}&=\frac{\Gamma_{ab}n_b}{m_a^2c^3}\frac{\thb\gamma }{u^3}\mu_{1},\\
\label{eq:Dperp}
D_{ab,\perp}&=\frac{\Gamma_{ab}n_b}{m_a^2c^3}\frac{1}{2\gamma u^3}\left(u^2(\mu_{0}+\gamma\thb\mu_{2})-\thb\mu_{1}\right).
\end{align}
The coefficients are illustrated in Fig.~\ref{fig:coefficients} as a function of $u$.

The special functions $\mu_0(u;\Theta)$, $\mu_1(u;\Theta)$, and $\mu_2(u;\Theta)$ are given by 
\begin{align}
\label{eq:mu0}
\mu_0&=\frac{\gamma^2L_0-\Theta L_1 + (\Theta-\gamma)u e^{(1-\gamma)/\Theta}}{e^{1/\Theta}\besselK_2(1/\Theta)},\\
\label{eq:mu1}
\mu_1&=\frac{\gamma^2L_1-\Theta L_0 + (\Theta\gamma-1)u e^{(1-\gamma)/\Theta}}{e^{1/\Theta}\besselK_2(1/\Theta)},\\
\label{eq:mu2}
\mu_2&=\frac{2\Theta\gamma L_1+(1+2\Theta^2)ue^{(1-\gamma)/\Theta}}{\Theta\; e^{1/\Theta}\besselK_2(1/\Theta)},
\end{align}
where the functions $L_0(u;\Theta)$ and $L_1(u;\Theta)$ are
\begin{align}
\label{eq:L0}
L_0& = \int_0^{u}ds\;\frac{e^{(1-\sqrt{1+s^2})/\Theta}}{\sqrt{1+s^2}},\\
\label{eq:L1}
L_1& = \int_0^{u}ds\;e^{(1-\sqrt{1+s^2})/\Theta}.
\end{align}
Our notation here differs from the Ref.~\cite{Pike_Rose_2014} as we have multiplied both the numerator and denominator in Eqs.~\eqref{eq:mu0} -~\eqref{eq:mu2} by $e^{1/\Theta}$ to avoid floating point errors when $\Theta$ is small.

The Monte Carlo simulation of a given test particle distribution is based on the connection between the Fokker-Planck equation and stochastic differential equations. 
If the distribution function $f_a(\bm{u},t)$ satisfies equation~\eqref{eq:fokker-planck}, then an individual sample particle from $f_a(\bm{u},t)$ obeys the following stochastic differential equation, known as the Langevin equation, of It\^o kind~\cite{coffey2004applications}
\begin{equation}
\label{eq:langevin equation for u_vec}
d\bm{u}(t)=\bm{K}_a(u(t),t)dt+\boldsymbol{\sigma}_a(u(t),t)\cdot d\bm{W},
\end{equation}
where the rank-2 tensor $\boldsymbol{\sigma}_a$ satisfies the condition
\begin{equation}
\label{eq:decomp 2sigmasigma=D}
\boldsymbol{\sigma}_a\boldsymbol{\sigma}_a^\intercal=2\mathbf{D}_a.
\end{equation}
Here $d\bm{W}$ is a differential of a vector-valued, uncorrelated standard Wiener processes $\bm{W}\sim{\cal N}(\bm{0},t\mathbf{I})$, with ${\cal N}$ being the standard multivariate normal distribution.
Since the diffusion tensor is diagonal, the decomposition Eq.~\eqref{eq:decomp 2sigmasigma=D} is easy to accomplish, and we find
\begin{equation}
\label{eq:diffusion tensor}
\boldsymbol{\sigma}_a=\sqrt{2 D_{a,\parallel}(u(t))}\bm{\hat{u}}\bm{\hat{u}}+\sqrt{2D_{a,\perp}(u(t))}\left(\mathbf{I}-\bm{\hat{u}}\bm{\hat{u}}\right).
\end{equation}

\section{Guiding-center test-particle operator}
\label{sec:gc}
In many applications the rapid oscillation of a charged particle in plane perpendicular to a magnetic field is of little interest.
If the magnetic field is slowly varying, one can resort to guiding center formalism which omits this gyro motion and depends only on the gyro-averaged quantities, thus reducing the 6D particle phase space into a 5D guiding center phase space.

Rigorous transformation of the particle phase-space Fokker-Planck equation~\eqref{eq:fokker-planck} into guiding-center phase-space was first carried out in Ref.~\cite{brizard2004guiding}. 
The transformation is based on the observation that the noncanonical particle phase-space Poisson bracket could be used to express the momentum-space derivatives according to
\begin{equation}
\frac{\partial\; \cdot\;}{\partial \mathbf{u}}=\{\mathbf{x},\;\cdot\;\},
\end{equation}
where $\mathbf{x}$ is the spatial coordinate.
This fact can be put to use by first observing that the isotropic test-particle diffusion tensor satisfies
\begin{equation}
\frac{\partial}{\partial\bm{u}}\cdot\mathbf{D}_{ab}=\frac{\partial D_{ab,\parallel}}{\partial u}\bm{\hat{u}},
\end{equation}
so that the particle phase-space collision operator can be written in a form
\begin{equation}
\frac{\partial f_a}{\partial t}=-\frac{\partial}{\partial\bm{u}}\cdot\left(\bm{Q}_af_a\right)+\frac{\partial}{\partial\bm{u}}\cdot\left(\mathbf{D}_a\cdot\frac{\partial f_a}{\partial\bm{u}}\right),
\end{equation}
where the modified friction coefficient is defined by
\begin{align}
\bm{Q}_a(\bm{u})&=\sum_b Q_{ab}(u)\bm{\hat{u}},\\
Q_{ab}(u)&=K_{ab}-\frac{\partial D_{ab,\parallel}}{\partial u}.
\end{align}
Now, using the Poisson brackets, one obtains
\begin{equation}
\frac{\partial f_a}{\partial t}=-\{x_i,Q_{a,i}f_a\}+\{x_i,D_{a,ij}\{x_j,f_a\}\},
\end{equation}
where the subscripts $i,j$ denote the Cartesian indices for the vector and tensor components, and summation over repeated indices is assumed. 

The guiding-center transformation then follows via Lie-transform (see Ref.~\cite{brizard2004guiding} for details), and the final form of the collision operator for the gyroangle-independent guiding-center distribution function $F$ becomes 
\begin{equation}
\frac{\partial F_a}{\partial t}=-\frac{1}{\gJ}\frac{\partial}{\partial \mathbf{Z}}\cdot\left(\gJ\boldsymbol{{\cal Q}}_{a} F_a+\gJ\boldsymbol{{\cal D}}_{a}\cdot\frac{\partial F_a}{\partial \mathbf{Z}}\right),
\end{equation}
where $\mathbf{Z}$ are guiding center phase-space coordinates and $\gJ$ is the transformation Jacobian.
The guiding center friction and diffusion coefficients are
\begin{align}
{\cal Q}_{a,\alpha}&=\langle \bm{\Delta}_{\alpha}\cdot {\cal T}^{-1}\bm{Q}_a\rangle\\
{\cal D}_{a,\alpha\beta}&=\langle \bm{\Delta}_{\alpha}\cdot {\cal T}^{-1}\mathbf{D}_a\cdot\bm{\Delta}_{\beta}\rangle,
\end{align}
where ${\cal T}^{-1}$ is the guiding-center push-forward, $\langle\cdot\rangle$ denotes a gyroaverage, and $\bm{\Delta}_{\alpha}$ are the so-called projection vectors, defined according to 
\begin{equation}
\bm{\Delta}_{\alpha}=\{{\cal T}^{-1}\bm{X},Z_{\alpha}\}_{gc},
\end{equation}
with $\{{\cal F},{\cal G}\}_{gc}$ being the guiding center Poisson bracket.
Here indices $\alpha,\beta$ denote the guiding center coordinates. 
Writing the guiding-center Fokker-Planck equation in a form similar to Eq.~\eqref{eq:fokker-planck}, we find
\begin{equation}
\label{eq:gc fokker-planck}
\frac{\partial F_a}{\partial t}=-\frac{1}{\gJ}\frac{\partial}{\partial \mathbf{Z}}\cdot\left(\gJ{\boldsymbol{\cal K}}_{a}F_a\right)+\frac{1}{\gJ}\frac{\partial}{\partial \mathbf{Z}}\frac{\partial}{\partial \mathbf{Z}}:\left(\gJ\boldsymbol{{\cal D}}_{a}F_a\right),
\end{equation}
where the drift coefficient is
\begin{equation}
\boldsymbol{{\cal K}}_{a}=\boldsymbol{{\cal Q}}_{a}+\frac{1}{\gJ}\frac{\partial}{\partial \mathbf{Z}}\cdot\left(\gJ\boldsymbol{{\cal D}}_{a}\right).
\end{equation}

The guiding center formalism rests on the assumption that the magnetic moment is invariant.
This invariance can be used to reduce the number of equations of motion to four by choosing the magnetic moment as one coordinate.
However, the diffusion tensor is not diagonal in this case~\cite{hirvijoki2013monte} but, as we saw in the last section, a diagonal basis is desired when considering the numerical implementation.
Fortunately, there exists a suitable set of coordinates where the diffusion tensor is (almost) diagonal.
This basis is $\gZv = (\mathbf{X},u,\xi)$ where $\mathbf{X}$ is the guiding center location, $u$ is the magnitude of the normalized momentum, and pitch is $\xi = \unit{u}\cdot\unit{b}$, where $\unit{b}$ is the unit vector parallel to the magnetic field $\mathbf{B}(\mathrm{X})$.
In these curvilinear coordinates the Jacobian is $\gJ = m_a B u$, and the diffusion tensor becomes
\begin{equation}
\mathbb{\gD}_{a} = \sum_b \gD_{ab,\mathbf{X}}(\mathbf{I}_\mathbf{X}-\unit{b}\unit{b}) + \sum_b \gD_{ab,u} \hat{u}\hat{u} + \sum_b \gD_{ab,\xi} \hat{\xi}\hat{\xi},
\end{equation}
with diagonal matrix $\mathbf{I}_\mathbf{X}$ having non-zero elements only in coordinates $\mathbf{X}$.
The friction coefficient has only one component
\begin{equation}
{\cal Q}_{ab,u} = -\frac{\Gamma_{ab}n_b}{m_a^2c^3}\frac{1}{u^2}\left(\frac{m_a}{m_b}\mu_{1}\right).
\end{equation}

The guiding center collision operator has few notable differences to the particle collision operator. 
First, the collisions now cause also spatial diffusion, with a diffusion coefficient~\cite{hirvijoki2013monte}
\begin{equation}
\gD_{ab,\mathbf{X}}=\left[(D_{ab,\parallel}-D_{ab,\perp})\frac{1-\xi^2}{2}+D_{ab,\perp}\right]\frac{c^2}{\Omega^2},
\end{equation}
where $\Omega = q_aB/m_a$ is the gyrofrequency.
In uniform magnetic field this corresponds to classical diffusion.
Second, the momentum magnitude and direction now have separate coefficients for the diffusion,
\begin{align}
\gD_{ab,u} &=2D_{ab,\parallel},\\ 
\gD_{ab,\xi} &= (1-\xi^2)\nu_{ab},
\end{align}
and for the drift
\begin{align}
{\cal K}_{ab,u} &= {\cal Q}_{ab,u}+D_{ab,\parallel}'+\frac{2D_{ab,\parallel}}{u},\\
{\cal K}_{ab,\xi}&=-\xi\nu_{ab},
\end{align}
where
\begin{equation}
\nu_{ab} = \frac{2D_{ab,\perp}}{u^2}
\end{equation} 
is the pitch collision frequency.
Third, both $\nu_{ab}$ and ${\cal K}_{ab,u}$ diverge at $u = 0$, which makes the particle collision operator more attractive when simulating thermal particles.
Also the guiding-center specific coefficients $\gD_{ab,\mathrm{X}}$, ${\cal K}_{ab,u}$, and $\nu_{ab}$ are illustrated in Fig.~\ref{fig:coefficients}.
Note that ${\cal K}_{ab,\mathrm{X}} = 0$, and that we have implicitly assumed uniform magnetic field when writing down the coefficients above, see Ref.~\cite{hirvijoki2013monte} for details.

The guiding center collision operator we have is in curvilinear coordinates so obtaining the corresponding Langevin equation is not as trivial as it was in the particle picture.
Details on the derivation of the Monte-Carlo operator are found in~\ref{app:curvilinear}, and here we only show the result
\begin{equation}
\label{eq: gc langevin}
d\gZv=\boldsymbol{{\cal K}}_{a}dt + \boldsymbol{\Sigma}_{a}\cdot d\W,
\end{equation}
where $\boldsymbol{\Sigma}_{a}$ is again easily obtained from the decomposition $(1/2)\boldsymbol{\Sigma}_{a}\boldsymbol{\Sigma}_{a}^\intercal = \boldsymbol{\gD}_{a}$.
Written explicitly, the guiding center collision operator is a set of equations
\begin{align}
\label{eq: SDE GC X}
d\mathbf{X} &= \sqrt{2\gD_{a,\mathrm{X}}}(\mathbf{I}-\unit{b}\unit{b})\cdot\W_\mathrm{X},\\
\label{eq: SDE GC u}
du &= {\cal K}_{a,u} dt + \sqrt{2D_{a,\parallel}}dW_u,\\
\label{eq: SDE GC xi}
d\xi &= -\xi\nu_a dt + \sqrt{(1-\xi^2)\nu_a}dW_\xi,
\end{align}
with $\W_\mathrm{X}$, $W_u$, and $W_\xi$ being independent Wiener processes.

\section{Monte Carlo algorithm}
\label{sec:MC operator}

The equation~\eqref{eq:langevin equation for u_vec} does not have a known analytical solution and, therefore, numerical methods are required.
Given initial condition $\bm{u}(t_0)$, the numerical approximation for $\bm{u}(t)$ is obtained by discretizing the time coordinate as $t_{k+1} = t_k + \Delta t$, $k= 0,\dots,n$, and evaluating $\bm{u}(t_{k+1})$ at each step using the solution of the previous step as an initial condition.
The evaluation is commonly done with the Euler-Maruyama method which, when leaving out the species subscript $a$ and switching to Einstein notation for brevity, reads
\begin{equation}
\label{eq:Euler Maruyama}
{u}_i(t_{k+1}) = {u}_i(t_{k}) + {K}_i(u(t_{k}))\Delta t + \sum_{j=1}^3 {G}_{ij}(u(t_{k}))\Delta W_j
\end{equation}
where the index $i=1,2,3$ represents the coordinates in Cartesian basis. 
In this basis, the matrix $\mathbf{G}$ reads
\begin{equation}
\label{eq:Euler Maruyama Gij}
G_{ij} = \sqrt{2D_\parallel}\hat{u}_i\delta_{ij}+\sqrt{2D_\perp}\left[\delta_{ij}-\hat{u}_i\hat{u}_j\right],
\end{equation}
where $\delta_{ij}$ is the Dirac delta.
The discretized differentials of Wiener processes are drawn from a normal distribution $\Delta W_i = W_i(t_{k+1})-W_i(t_{k})\sim {\cal N}(0,\Delta t)$.

Implementing Euler-Maruyama method is straightforward, and all the struggle is in evaluating the coefficients $\mathbf{K}$ and $\mathbf{D}$ that, through the special functions $\mu_0$, $\mu_1$, and $\mu_2$, depend on the integrals $L_0$ and $L_1$. 
These integrals (Eqs.~\eqref{eq:L0} and~\eqref{eq:L1}) cannot be solved analytically and, therefore, we evaluate them via the adaptive Simpson's method. 
The adaptive Simpson's method divides the integration interval into subintervals until the difference in the resulting numerical approximation between the successive divisions is less than a given tolerance.
However, most of the contribution to $L_0$ and $L_1$ comes from small values of $u$, and the adaptive Simpson's method converges to an incorrect value if the upper limit for $u$ is large while $\Theta$ is small.
This can be avoided by noting that both integrands equal to unity at $u=0$ and, as the integrands decay rapidly, the integrals have practically constant values beyond a certain point.
Therefore, we can make the integration robust by defining a cut-off limit for $u$ as
\begin{equation}
\label{eq:cutoff limit}
e^{(1-\sqrt{1+u_c^2})/\Theta} = \epsilon\;\Rightarrow\; u_c = \sqrt{(1-\Theta \ln\epsilon)^2-1},
\end{equation}
where the accuracy is controlled with parameter $\epsilon$, and then applying the adaptive Simpson's method separately on intervals $[0,u_c]$ and $[u_c,u]$ (when $u_c < u$). 

\begin{figure}[!t]
\centering
\begin{overpic}[width=0.45\textwidth]{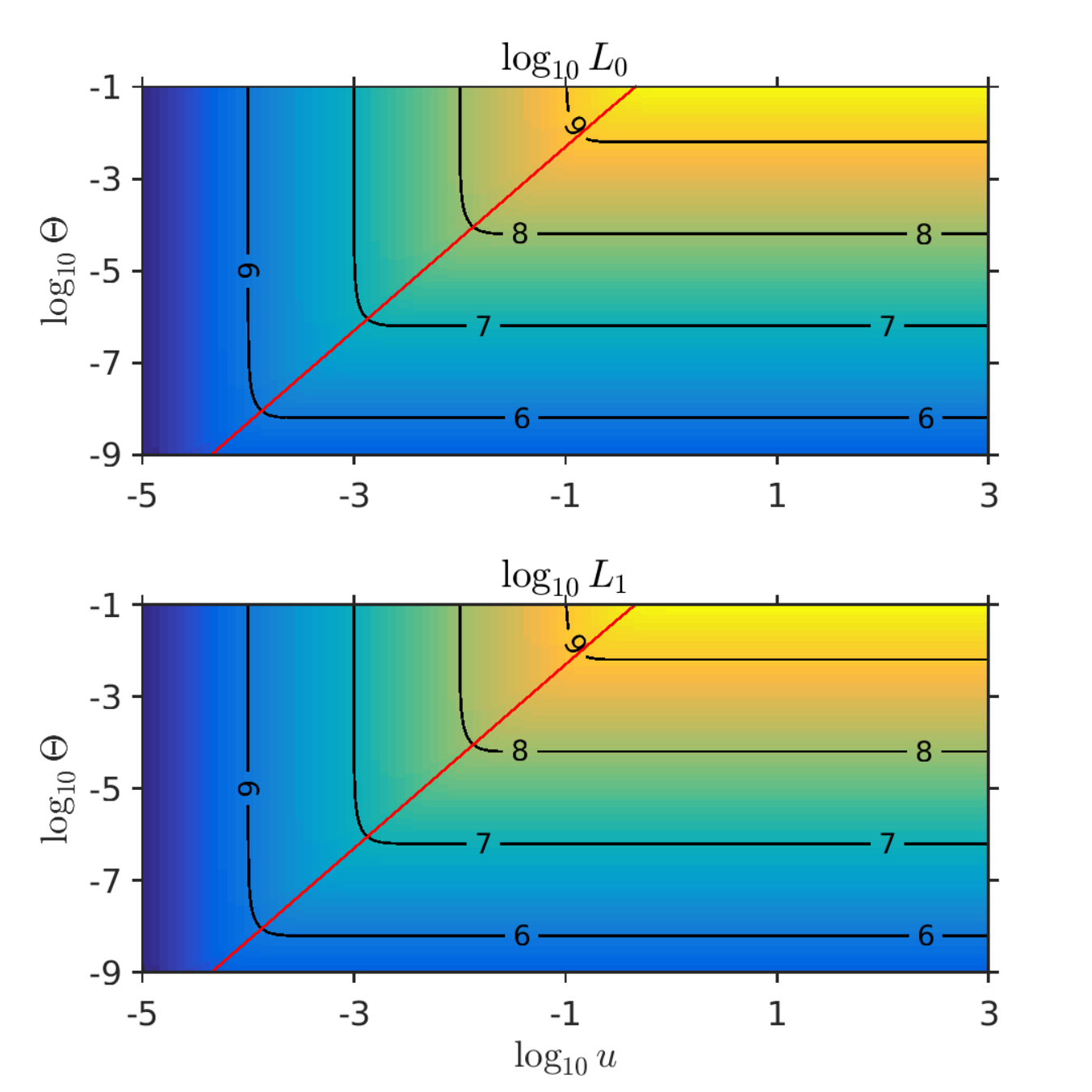}
\put(2,95){a)}
\put(2,45){b)}
\end{overpic}
\caption{
Integrals~\ref{eq:L0} and~\ref{eq:L1} as a function of $u$ and $\Theta$. Black lines are contours of the respective integrals whereas red line is the curve $u=\sqrt{\Theta^2+2\Theta}$. The values tabulated for this figure are enough to cover most of the tokamak plasmas as, for example, a slow 10 eV tungsten impurity has $u \approx 1\times 10^{-5}$, a fast 1 GeV runaway electron has $u \approx 2\times 10^3$ while cold 10 eV helium have $\Theta \approx 3\times10^{-9}$, and hot 50 keV electrons have $\Theta \approx 1\times10^{-1}$.
Note that the integrals have very similar values.
} 
\label{fig:L0L1}
\end{figure}

However, evaluating the integrals this way at every integration time step would be very inefficient.
Therefore, we calculate and tabulate the $L_0$ and $L_1$ values for a wide range of $u$ and $\Theta$ values before proceeding to solve Eq.~\eqref{eq:Euler Maruyama}.
From the cut-off limit, Eq.~\eqref{eq:cutoff limit}, we can deduce that the integrals are sensitive to $u$ and $\Theta$ only near the curve $u=\sqrt{\Theta^2+2\Theta}$, but not elsewhere as Fig.~\ref{fig:L0L1} illustrates.
This means that $L_0$ and $L_1$ do not have to be tabulated with high resolution and, therefore, the tabulated values can be used to efficiently compute the coefficients $\mathbf{K}$ and $\mathbf{D}$ at every time step.

\section{Integration with an adaptive time step}
\label{sec:adaptive}

The collision operator introduced in the last section, Eq.~\eqref{eq:Euler Maruyama}, was based on the Euler-Maruyama method and, therefore, is not applicable for an adaptive time step scheme.
An adaptive integration is desirable since especially the pitch collision frequency has a strong dependency on test particle momentum (recall Fig.~\ref{fig:coefficients}).
For example, in a proton-electron plasma with $T = 10$ keV, $n=10^{20}$ m$^{-3}$, a thermal electron has a pitch collision frequency in the order of $\nu \approx 10^3$  $s^{-1}$, while, for a high-energy electron with $E_\mathrm{kin} = 1$ MeV, the frequency is two orders of magnitude less $\nu \approx 10$ $s^{-1}$.

The simplest method which is suitable for the adaptive scheme is the Milstein method which has both weak and strong convergence of 1.0.
When the evolution of an $N$-dimensional stochastic variable $\Y$ is given by the following It\^o form
\begin{equation}
\label{eq:Langevin equation}
d\Y = \mathbf{p}(\Y)dt + \mathbf{g}(\Y)d\W,
\end{equation}
the Euler-Maruyama method is acquired by simple discretization: $d\Y\rightarrow\Delta \Y$, $dt\rightarrow\Delta t$, and $d\W\rightarrow\Delta \W$.
The Milstein method 
\begin{multline}
\label{eq:Milstein method general}
\y_i(t_{k+1}) = \y_i(t_{k}) + p_i(\y(t_{k}))\Delta t 
+ \sum_{j=1}^N g_{ij}(\y(t_{k}))\Delta W_j \\
+\frac{1}{2}\sum_{j=1}^N\sum_{n=1}^N g_{nj}(\y(t_{k}))\frac{\partial}{\partial\y_n}g_{ij}(\y(t_{k})) I_{nj}(t_k),
\end{multline}
has a form similar to the Euler-Maruyama, with the difference being the additional last term, which contains the double It\^o integral
\begin{equation}
I_{ij}(t_k) = \int_{t_k}^{t_{k+1}} dW_i(t)dW_j(t).
\end{equation}

In the adaptive scheme, the integration time step is varied so that local truncation error stays within user-defined boundaries.
For our purposes, we follow the scheme proposed in Ref.~\cite{lamba2003adaptive}, which was later refined in Ref.~\citep{lambastepsize}, in which two different error controls are used.
The first error control, defined in terms of the deterministic component $\mathbf{p}$, is
\begin{equation}
\label{eq:edrift}
\edrift = \max_{i = 1,\dots,N}\left\{\frac{1}{2\eabs{i}}\left| p_i\frac{\partial p_i}{\partial \y_i}(\Delta t)^2\right| \right\},
\end{equation}
where $\eabs{i}$ is the tolerated error for the variable $Y_i$.
The second error control is defined in terms of the stochastic component $\mathbf{g}$,
\begin{equation}
\label{eq:ediff1}
\ediff{1} = \max_{i = 1,\dots,N}\left\{\frac{1}{6\eabs{i}}  \left|g_{ii}\left(\frac{\partial g_{ii}}{\partial \y_i}\right)^2(\Delta W_i)^3\right| \right\}.
\end{equation}
Alternatively, the second error control can be defined in terms of both $\mathbf{p}$ and $\mathbf{g}$
\begin{equation}
\label{eq:ediff2}
\ediff{2} = \max_{i = 1,\dots,N}\left\{\frac{1}{6\eabs{i}}  \left| g_{ii}\frac{\partial p_i}{\partial \y_i}J_i\right| \right\},
\end{equation}
which is useful when $\mathbf{g}'$ is zero or expensive to compute.
Here $J$ is the Stratonovich integral, $J_i = \int_{t_k}^{t_{k+1}}\int_{t_k}^t\circ dW_idt$, which has the value~\cite{lambastepsize}
\begin{equation}
J_i = \frac{1}{2}\Delta t \left(\Delta W_i + \frac{\omega_i}{\sqrt{3}}\right),
\end{equation}
where $\omega_i\sim \mathcal{N}(0,\Delta t)$. 
Approximating $|\omega_i|\approx\sqrt{\Delta t}$, we get $|J_i| = (1/2)\Delta t(|\Delta W_i| + \sqrt{\Delta t/3})$.
One can see that the error controls given by Eqs.~\eqref{eq:ediff1} and~\eqref{eq:ediff2} are of the order $\mathcal{O}((\Delta t)^{3/2})$, while Eq.~\eqref{eq:edrift} has higher order $\mathcal{O}((\Delta t)^2)$. 
The latter then seems unnecessary. 
However, this is not the case in the limit of weak diffusion, where $\mathbf{g}\approx 0$, and the deterministic component $\mathbf{p}$ dominates.

\subsection{Particle operator}

The Milstein method for the particle operator is obtained from by first substituting the tensor $\mathbf{G}$, Eq.~\eqref{eq:Euler Maruyama Gij}, in the extra term in Eq.~\eqref{eq:Milstein method general}.
The next task would be to discretize the double integral $\mathbf{I}$, but this turns out to be fatal to our approach.
The diagonal components of $\mathbf{I}$ can be shown to be exactly $I_{ii} = (\Delta W_i)^2-\Delta t$, but the non-diagonal ones have to be approximated~\cite{malham2010introduction,ryden2001simulation}.
There are ways to do this but they come with a hefty price: the strong order of converge would be reduced to 0.5.
Since we chose the Milstein method for the exact reason that its strong order is 1.0, this is obviously unacceptable.
On the other hand, approximations that would maintain the strong order 1.0 are complicated and expensive to compute, reducing the efficiency gains of the adaptive integration, so we seek a better way to implement the operator.

The fact that $\mathbf{G}$ is diagonal in $(u_\parallel,u_\perp)$ basis provides us with a suitable path.
With only diagonal components included, i.e. the noise being commutative, the Milstein method reads
\begin{multline}
\y_i(t_{k+1}) = \y_i(t_{k}) + p_i(\y(t_{k}))\Delta t 
+ g_{ii}(\y(t_{k}))\Delta W_i \\
+\frac{1}{2} g_{ii}(\y(t_{k}))\frac{\partial}{\partial\y_i}g_{ii}(\y(t_{k})) \left[(\Delta W_i)^2-\Delta t\right],
\end{multline}
so that the collision operator becomes
\begin{align}
\label{eq:milstein uperp}
u_{\perp,j}(t_{k+1}) &= \sqrt{2D_\perp(u(t_k))}\Delta W_j,\;j=1,2, \\
\label{eq:milstein upar}
u_\parallel(t_{k+1}) &= K(u(t_k))\Delta t + \sqrt{2D_\parallel(u(t_k))}\Delta W_3 \nonumber \\
&+ \frac{1}{2}D_\parallel'(u(t_k))[(\Delta W_3)^2-\Delta t], \\
\label{eq:milstein utot}
u_i(t_{k+1}) &= u_i(t_{k}) + u_\parallel(t_{k+1})\hat{u}_i + \sum_{j=1}^2 u_{\perp,j}(t_{k+1})\hat{\perp}_j
\end{align}
where $(\unit{u},\unit{\perp}_1,\unit{\perp}_2)$ form an orthogonal basis, and prime denotes partial derivative with respect to $u$.
Equation~\eqref{eq:milstein uperp} has formally reduced to the Euler-Maruyama method as the noise is additive, i.e., $\partial D_\perp/\partial u_\perp = 0$, so this is a special case where even the Euler-Maruyama method has strong convergence of 1.0.
The error terms, Eqs.~\eqref{eq:edrift} and~\eqref{eq:ediff1}, are now solely determined by Eq.~\eqref{eq:milstein upar}:
\begin{align}
\eabs{} &= \etol (|K\Delta t|+\sqrt{2D_\parallel\Delta t}),\\
\edrift &= \frac{\left|KK'\right|(\Delta t)^2}{2\eabs{}},\\
\label{eq:milstein prt ediff1}
\varepsilon_\mathrm{diff} &= \frac{(D_\parallel')^2|\Delta W_3|^3}{6\eabs{}\sqrt{D_\parallel}}.
\end{align}
where we have chosen Eq.~\eqref{eq:ediff1} as the second error estimate instead of Eq.~\eqref{eq:ediff2}. 
This is because $K'$ can be zero (recall Fig.~\ref{fig:coefficients}), in which case both error estimates, Eqs.~\eqref{eq:edrift} and~\eqref{eq:ediff2}, would yield zero values.

The Milstein method requires additional computation of $D_\parallel'$, while $K'$ is needed for the error estimates. 
Both can be evaluated analytically ($D_\perp'$ is included for completeness sake):
\begin{align}
\frac{\partial K}{\partial u}&=\frac{\Gamma_{ab} n_b}{m_a^2c^3}\frac{1}{u^3}\left[ 2\left(\frac{\mu_0}{\gamma}+\frac{m_a}{m_b}\mu_1\right)\right.\nonumber
\\
&\left.-u\left(\frac{\mu_0'}{\gamma} +\frac{m_a}{m_b}\mu_1'\right)+u^2\frac{\mu_0}{\gamma^3}\right],
\\
\frac{\partial D_\parallel}{\partial u}&=\frac{\Gamma_{ab} n_b}{m_a^2c^3}\frac{\Theta_b }{\gamma u^4}
\left[u\gamma^2 \mu_1'- (1+2\gamma^2)\mu_1\right],
\\
\frac{\partial D_\perp}{\partial u}&=\frac{\Gamma_{ab} n_b}{m_a^2c^3}\frac{1}{2 \gamma^3 u^4}\left[(4\gamma^2-1)\Theta_b \mu_1 -u\Theta_b \gamma^2\mu_1'\right.\nonumber
\\
&\left.-u^2((2\gamma^2-1)\mu_0+\Theta_b\gamma^3\mu_2) + u^3\gamma^2(\mu_0'+\Theta_b\gamma\mu_2')\right],
\end{align}
where the derivatives of the special functions are
\begin{align}
&\mu_0'=\frac{ 2 \Theta\gamma u L_0 +(\gamma -2\Theta )u^2 e^{(1-\gamma)/\Theta }  }{ \Theta\gamma e^{1/\Theta }K_2\left(\frac{1}{\Theta }\right)},\\
&\mu_1'=\frac{2 \Theta\gamma u  L_1 + \left(1+2 \Theta^2\right)u^2e^{(1-\gamma)/\Theta} }{\Theta\gamma   e^{1/\Theta }K_2\left(\frac{1}{\Theta }\right)},\\
&\mu_2'=\frac{2\Theta^2 u L_1  +  \left(2\Theta^3\gamma+2\Theta^2 +\Theta\gamma-u^2\right)e^{(1-\gamma)/\Theta }}{\Theta ^2\gamma e^{1/\Theta }K_2\left(\frac{1}{\Theta }\right)}.
\end{align}
One can note that $\mu_1'=(u/\gamma)\mu_2$.

\subsection{Guiding center operator}

With Milstein discretization, the guiding center collision operator, Eqs.~\eqref{eq: SDE GC X}~-~\eqref{eq: SDE GC xi}, becomes
\begin{align}
\label{eq:GC X}
X_i(t_{k+1}) &= X_i(t_{k})+\sum_{j=1}^3\sqrt{2\gD_{\mathbf{X},ij}(\gZv(t_k))}\left(\delta_{ij}-\hat{b}_i\hat{b}_j\right)\Delta W_{\mathrm{X},j},\\
\label{eq:GC u}
u(t_{k+1}) &= u(t_{k}) + \mathcal{K}_u(\gZv(t_k))\Delta t + \sqrt{2D_\parallel(\gZv(t_k))}\Delta W_u \nonumber \\
&+ \frac{1}{2}D_\parallel'(\gZv(t_k))\left[(\Delta W_u)^2-\Delta t\right],\\
\xi(t_{k+1}) &= \xi(t_{k}) - \xi(t_{k})\nu(\gZv(t_k))\Delta t \nonumber \\
&+ \sqrt{(1-\xi^2(t_{k}))\nu(\gZv(t_k))}\Delta W_\xi \nonumber \\
\label{eq:GC xi}
&- \frac{1}{2}\xi(t_{k})\nu(\gZv(t_k))\left[(\Delta W_\xi)^2-\Delta t\right].
\end{align}
Equation~\eqref{eq:GC X} was reduced to the Euler-Maruyama form as we have assumed uniform magnetic field, i.e., $(\partial/\partial\mathrm{X})(\cdot) = 0$.
In addition to these equations, there are also boundary conditions.
The particle pitch is limited to the interval $\left[-1,1\right]$ which can be enforced with a reflecting boundary condition,
\begin{equation}
\xi(t_{k+1}) = \sign(\xi(t_{k+1}))(2-|\xi(t_{k+1})|),
\end{equation}
applied if $|\xi(t_{k+1})| > 1$.
Similarly, $u$ cannot have negative values, so a reflecting boundary condition should be set at $u = 0$.
However, as both $\nu$ and $\mathcal{K}_u$ diverge at $u = 0$ (recall Fig.~\ref{fig:coefficients}), setting the reflecting boundary condition to a small positive value but still below the thermal momentum value $u\approx\sqrt{2\Theta}$, e.g. at $u= 0.05\sqrt{2\Theta}$, ensures that the time step in the adaptive scheme does not become extremely small, and that the method is stable if a fixed time step is used.

The error estimates from Eqs.~\eqref{eq:edrift}~-~\eqref{eq:ediff2} are
\begin{align}
\eabs{u} &= \etol \left(|\mathcal{K}_u|\Delta t + \sqrt{2D_\parallel \Delta t}\right),\\
\label{eq:gc drift error}
\edrift &= \max\left\{\frac{\left|\mathcal{Q}_u\mathcal{Q}_u'\right|}{2\eabs{u}}
,\frac{\left|\xi\nu^2\right|}{2\eabs{\xi}}\right\}(\Delta t)^2,\\
\label{eq:gc diff error}
\varepsilon_\mathrm{diff} &= \max\left\{\frac{\left|D_\parallel^2 (\Delta W_u)^3\right|}{6\eabs{u}\sqrt{D_\parallel}},\right.\nonumber\\
&\left. \frac{\sqrt{1-\xi^2}\nu^{3/2}\left|\Delta W_\xi+\sqrt{\Delta t/3})\right|\Delta t}{2 \eabs{\xi}}\right\}.
\end{align}
Note that we have two tolerances: $\eabs{u}$ for the momentum and $\eabs{\xi}$ for the pitch which we set $\eabs{\xi} = \etol$ as the pitch values are bounded in an interval.
Strictly speaking, the error tolerance Eq.~\eqref{eq:gc drift error}, should be defined in terms of $\mathcal{K}_u$, but using $\mathcal{Q}_u$ we only have to evaluate the derivative 
\begin{equation}
\frac{\partial\mathcal{Q}_{ab,u}}{\partial u} = -\frac{\Gamma_{ab}n_b}{m_a^2c^3}\frac{1}{u^3}\frac{m_a}{m_b}(u \mu_1'-\mu_1).
\end{equation}
$\mathcal{Q}_u$ is the dominant term in $\mathcal{K}_u$ outside the diffusion dominated regime so this alteration is justifiable.
For the $\xi$ error term in Eq.~\eqref{eq:gc diff error}, we had to choose diffusion error estimate Eq.~\eqref{eq:ediff2} instead because the estimate derived from Eq.~\eqref{eq:ediff1}
\begin{equation}
\sqrt{2\gD_\xi}\left(\frac{\partial \sqrt{2\gD_\xi}}{\partial \xi}\right)^2= \frac{\xi^2}{\sqrt{1-\xi^2}}\nu^{3/2},
\end{equation}
diverges when $|\xi| = 1$.
No separate error limit was set for the spatial coordinate $\mathrm{X}$.

\subsection{Optimal time step and Brownian bridge}

Now that the criteria for time step rejection are established, the next task is to choose an optimal time step to minimize the number of rejections.
A good guess for the initial step is $\Delta t_\mathrm{init} = \epsilon_\mathrm{tol}^{3/2}/\nu$, but the presence of $|\Delta\W|$ in the error estimates complicates finding the optimal time step. 
The simplest scheme, known as the Brownian tree, is based on halving and doubling the current time step. 
However, this scheme is far from optimal and, therefore, we choose to implement the algorithm, described in detail in Ref~\cite{lamba2003adaptive}, where the next time step depends on the value of the error estimates. 
In the regime of weak diffusion, $\edrift > \varepsilon_\mathrm{diff}$, we could treat the collision operator as an ODE, and choose the next time step as
\begin{equation}
\Delta t' = \min(1.5,\beta \edrift^{-1/2})\Delta t,
\end{equation}
where $\beta < 1$ is a safety factor for which we set value $\beta = 0.9$.
However, $\edrift > \varepsilon_\mathrm{diff}$ could also be due to extraordinary small $|\Delta W|$, not because drift dominates, so the next time step is chosen as
\begin{equation}
\Delta t_\mathrm{next} = n\frac{\Delta t'}{3}, 
\end{equation}
where
\begin{equation}
n = \max\{l:\;|\Delta W_n| < \Delta W_\mathrm{opt}, \; \forall l= 1,\dots, 3\}.
\end{equation}
Here $\Delta W_\mathrm{opt} = \beta\varepsilon_\mathrm{diff}^{-1/3}|\Delta W|$ is the estimate for the ``optimal'' value of Wiener process.
In the diffusion dominated regime, $\varepsilon_\mathrm{diff} > \edrift $, we again determine the next step iteratively from
\begin{equation}
\Delta t_\mathrm{next} = n\frac{\Delta t}{3}, 
\end{equation}
where $n$ is determined by the condition
\begin{equation}
n = \max\{l:\;|\Delta W_n| < \Delta W_\mathrm{opt}, \; \forall l= 1,\dots, l_\mathrm{max}\},
\end{equation}
where $l_\mathrm{max} = 2$ if the current step was rejected, $l_\mathrm{max} = 4$ if $\Delta W /\sqrt{\Delta t} < 2$, i.e., the current Wiener process value was not an outlier, and otherwise $l_\mathrm{max} = 6$.

Whenever new Wiener processes are generated, these must always be stored -- even when the time step they are associated with is rejected or they are generated for the sole purpose of determining the next time step.
Only when integration has reached time $t$ can processes $\W(t')$, $t' <t$, be discarded.
The reason for this is that the realized values condition the distribution of the Wiener processes, so discarding them can lead to a bias if the discard mechanism is systematic.
A systematic mechanism can arise, e.g., from error estimate Eq.~\eqref{eq:ediff1} as discarding Wiener processes would lead to over-representation of small values of $\Delta W$ when $\mathbf{g}$ and $\mathbf{g}'$ are large.

The bias is avoided by introducing the conditioned probability distribution known as the Brownian bridge. 
First, let $\W_{\tm}$ and $\W_{\tp}$ be adjacent realized Wiener processes with $\tm < \tp$. 
When $\tm < t < \tp$, the process $\W_t$ is no longer normally distributed as $\mathcal{N}(\W_{\tm},(t-\tm)\mathbf{I})$, but follows a different normal distribution where the mean is
\begin{equation}
\label{eq:brownian bridge mean}
\boldsymbol{\mu} = \W_{\tm} + ( \W_{\tp}-\W_{\tm} )\frac{t-\tm}{\tp-\tm},
\end{equation}
and the variance
\begin{equation}
\label{eq:brownian bridge var}
\Sigma = \frac{(t-\tm)(\tp-t)}{\tp-\tm}.
\end{equation}
One can observe that the variance of $\W_t$ has its maxima at the center of the interval $[\tm,\tp]$ while the expected values follow a straight line from $\W_{\tm}$ to $\W_{\tp}$. 
Only when $\W_{\tp}$ for $\tp > t$ does not exist, $\W_t\sim\mathcal{N}(\W_{\tm},(t-\tm)\mathbf{I})$.

\section{Verification and benchmark}
\label{sec:verification}

To summarize, we have now developed following collision operators: the fixed step Euler-Maruyama method in the particle phase space, Eq.~\eqref{eq:Euler Maruyama}, the adaptive Milstein method in the particle phase space, Eqs.~\eqref{eq:milstein uperp}~--~\eqref{eq:milstein utot}, and the adaptive Milstein method in the guiding center phase space, Eqs.~\eqref{eq:GC X}~--~\eqref{eq:GC xi}. From now on, we refer to these as \FEP, \AMP, and \AMG, respectively.
These operators should yield equivalent results which preferably are the same as those obtained analytically -- a topic we investigate here. 

\begin{figure}[t]
\centering
\begin{overpic}[width=0.45\textwidth]{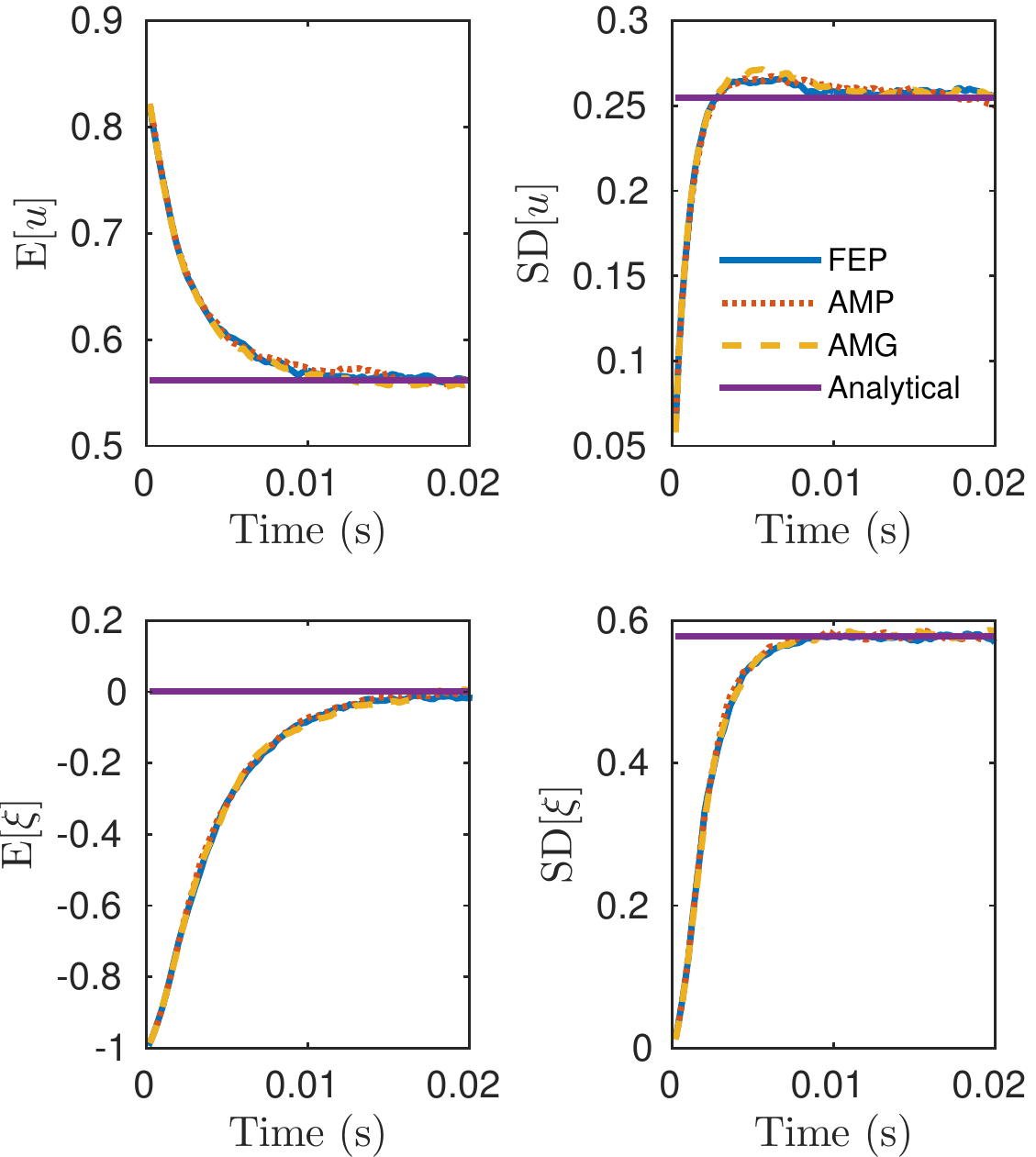}
\put(0,100){a)}
\put(47,100){b)}
\put(0,50){c)}
\put(47,50){d)}
\end{overpic}
\caption{Time evolution of the numerical solution of the particle distribution computed with different methods (\FEP{}, \AMP{}, and \AMG{}) and the thermal equilibrium (Analytical). (a) Mean and (b) standard deviation of the momentum distribution. (c) Mean and (d) standard deviation of the pitch distribution. Initially each test particle had $u=\sqrt{(1+3\Theta)^2-1}$ and $\xi=-1$.
}
\label{fig:time evolution}
\end{figure}

Our first task is to verify that a given test particle population relaxes to Maxwell-J\"uttner distribution, Eq.~\eqref{eq:maxwell juttner}. 
In equilibrium, the magnitude of the momentum is distributed as
\begin{equation}
\label{eq:u dist}
u \sim u^2e^{-\sqrt{1+u^2}/\Theta},
\end{equation}
and the pitch is distributed as $\xi\sim\mathcal{U}(-1,1)$, where $\mathcal{U}$ is the uniform distribution. 
According to Eq.~\eqref{eq:u dist}, the test particle mean momentum should converge to $0.56$ and variance to $3.62$
when considering a test case where both background and the test particle population consist of (relativistic)  electrons with $\Theta = 1\times10^{-1}$.  
Likewise mean pitch should converge to $0$ and variance to $1/3$.
From Fig.~\ref{fig:time evolution} we see that this is the case: All operators converge to the equilibrium values at the same rate and, thus, are verified in this regard.

The above test did not verify the spatial collision operator, Eq.~\eqref{eq:GC X}, which, in a uniform magnetic field, should correspond to the classical diffusion given by the coefficient
\begin{equation}
\label{eq:classical diffusion}
D_c = \int_0^\infty du\frac{1}{2}\rho^2(u)\nu(u)f(u),
\end{equation}
where $\rho$ is the Larmor radius, $\nu$ is the pitch collision frequency, and $f(u)$ is the momentum distribution function.
The spatial diffusion coefficient $D_B$ can be estimated with a Monte Carlo method as $D_B = \sum_{j=1}^n D_j/n$, where $D_j$ are test particle diffusion coefficients, and $n$ is either the number of test particles or, for a single particle, the number of time steps.
It can be shown that the ratio $D_j/D_X$ obeys $\chi^2$ distribution with 1 degree of freedom~\cite{boozer1981monte}.
With the \AMG{} operator, $D_j$ is easy to calculate as $D_j = (\Delta X_j)^2/2\Delta t_j$, where $\Delta X$ is the change in guiding center position, along some predefined direction, during time $\Delta t$.
With the \AMP{} or \FEP{} operator, the collisions only affect particle momentum and, hence, do not lead to spatial diffusion, unless the particle collision operator is coupled with the Lorentz force.
For numerical evaluation of the Lorentz force, we use the energy conserving scheme~\cite{zhang2015volume}.

\begin{figure}[t]
\centering
\begin{overpic}[width=0.45\textwidth]{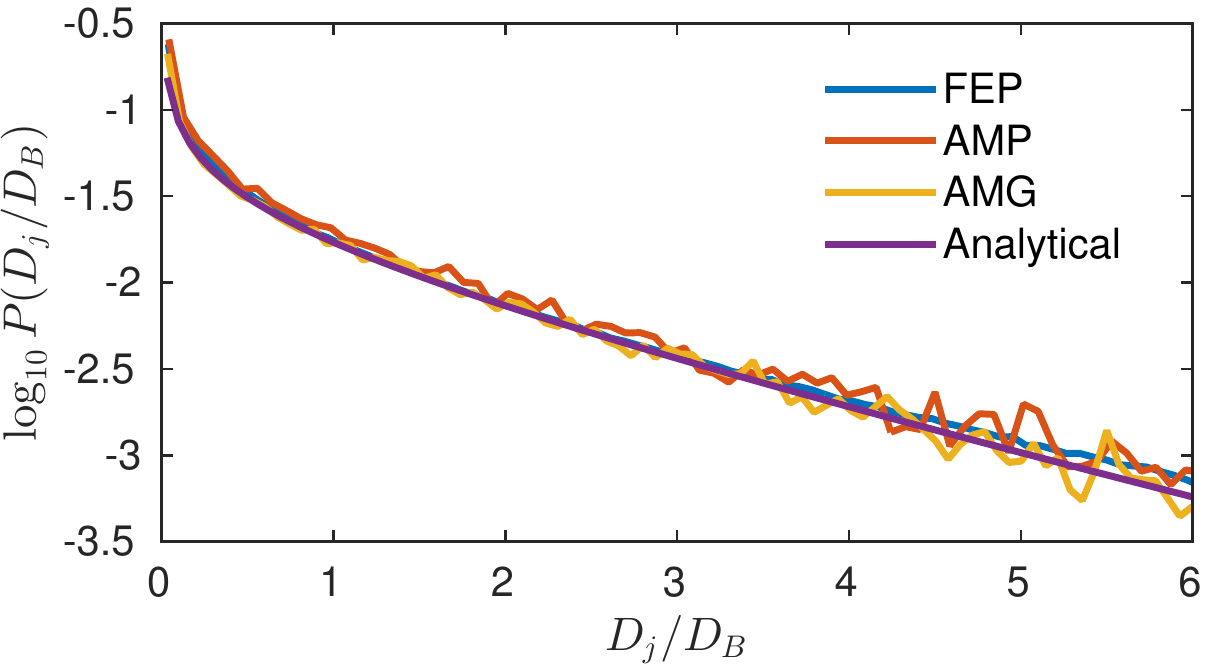}
\end{overpic}
\caption{Diffusion coefficient distribution, $P(D_j/D_B)$, when evaluated with a Monte Carlo method using different collision operators.
The analytical distribution is $\chi^2_1(D_j/D_B)$ where $D_B=D_c$ is the classical diffusion coefficient obtained from Eq.~\eqref{eq:classical diffusion}.
The particle diffusion depends not only on the background plasma but also on magnetic background, which was here uniform with a magnitude $B = 5$ T.
}
\label{fig:diffusion coefficient}
\end{figure}

Using a test case where the test particles are thermalized, and evaluating diffusion coefficient with different operators, we obtain the test particle diffusion coefficient distributions shown in Fig.~\ref{fig:diffusion coefficient}. 
We thus confirm that all operators yield results matching well to the analytical result calculated directly from Eq.~\eqref{eq:classical diffusion}.

We have not yet shown that the rate at which test population relaxes to equilibrium is correct, or that the operators are valid also in the regime of low diffusivity, where the deterministic parts in the collision operators dominate.
To tackle both issues, we perform a slowing-down simulation using the same background plasma but this time with test particles having initially $u=5$.
In the slowing-down simulation, fast test particles are simulated until they reach a certain momentum below which they can be considered part of the thermal population.
Here the common figure of merit is the slowing-down time, i.e., the time it takes for a particle to thermalize.

\begin{figure}[t]
\centering
\begin{overpic}[width=0.45\textwidth]{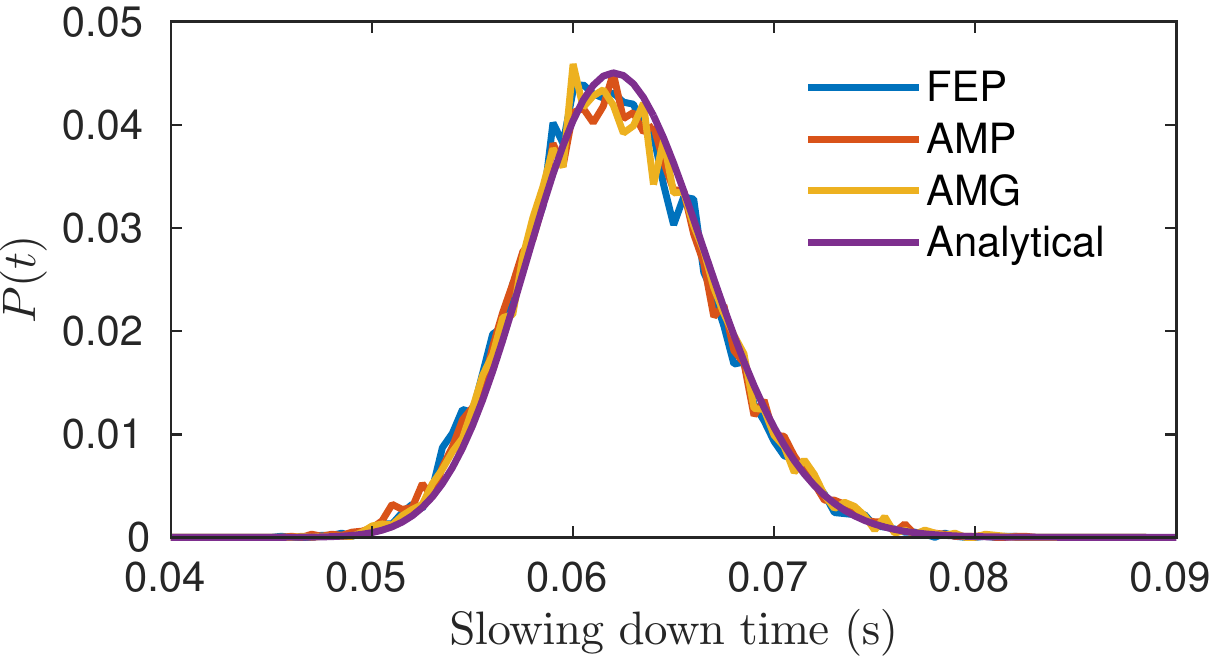}
\end{overpic}
\caption{Verification that the developed numerical numerical methods yield a slowing-down time distribution, $P(t)$, similar to the analytical estimate.
The analytical estimate corresponds to inverse Gaussian distribution $IG(\alpha/\mathcal{K}_u,\alpha^2/(2D_\parallel))$ where we have chosen parameters  $\alpha = -4$, $\mathcal{K}_u=-64\;\mathrm{s}^{-1}$, and $D_\parallel=0.65\;\mathrm{s}^{-1}$.
}
\label{fig:slowing down time}
\end{figure}

There are ways to analytically estimate the slowing-down time~\cite{anderson2004slowing}, but here we derive a simple estimate from the theory of stochastic processes.
The slowing-down process is analogous to the problem of finding the so-called first passage time. 
Consider a stochastic 1D process defined by the Langevin equation~\eqref{eq:Langevin equation} where coefficients $p$ and $g$ are constant.
If this process has an initial value $Y(t_0)=Y_0$, then the time, $\tau$, it takes to reach value $Y_0+\alpha$ for the first time, obeys an inverse Gaussian distribution $\tau\sim IG(\alpha/p,\alpha^2/g^2)$.
Here the 1D process in question is the guiding center momentum equation~\eqref{eq:GC u}, $u_{0}$ is the particle initial momentum and $u_\mathrm{min}$ is the momentum value below which a particle can be considered thermalized.
Setting $u_{0} = 5$ and $u_{min}  =1$, we have $\alpha = -4$.
Furthermore, setting the background temperature to $\Theta = 1\times10^{-2}$, the coefficients $Q$ and $D_\parallel$ are approximately constant on the interval $[u_{min},u_{0}]$ (recall Fig.~\ref{fig:coefficients}).
Figure~\ref{fig:slowing down time} shows the slowing down distribution estimated this way overlapping with the distributions obtained numerically with the different collision operators.
We can therefore conclude that the operators yield a correct relaxation rate and are valid also in the regime of low diffusionality.

\begin{figure}[t]
\centering
\begin{overpic}[width=0.45\textwidth]{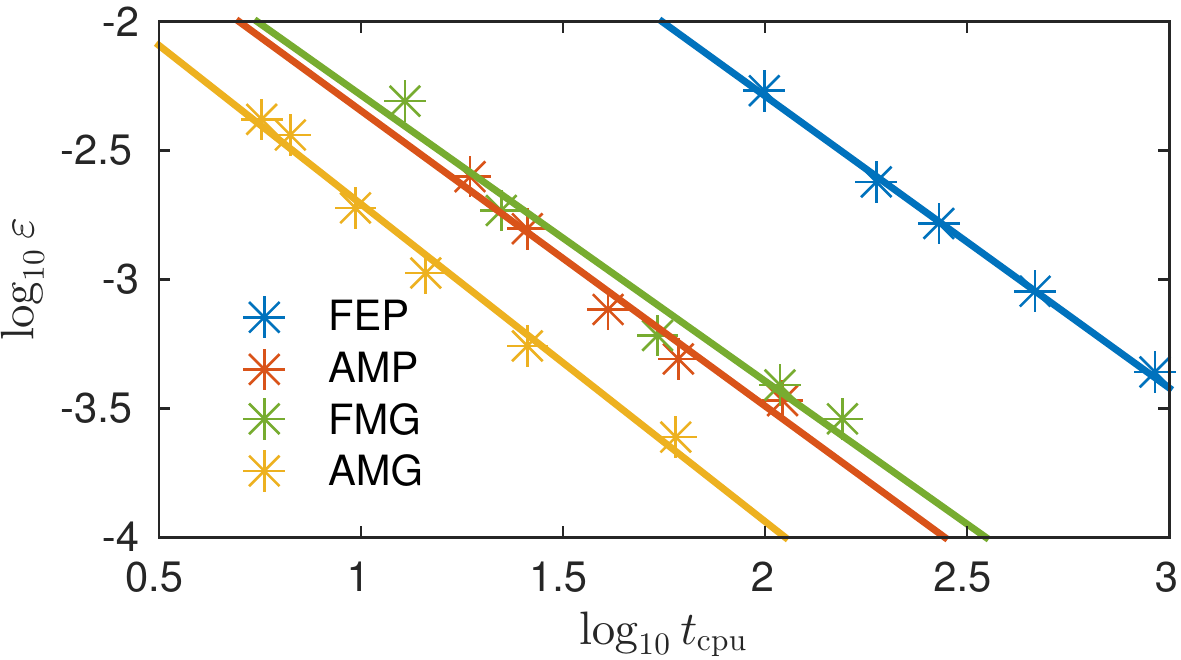}
\end{overpic}
\caption{Results of the benchmark between fixed (\FEP{} and \FMG{}) and adaptive (\AMP{} and \AMG{}) time step methods.
Each marker correspond to a test case, and they show the relative error, $\epsilon$, in the computed slowing down time as a function of the required computational time $t_\mathrm{cpu}$. 
$\epsilon$ decreases while $t_\mathrm{cpu}$ increases when error tolerances are tightened between subsequent test cases.
The general trend for each operator is illustrated with a fitted line.
}
\label{fig:benchmark}
\end{figure}

Having verified the collision operators, it is time to benchmark them to see what can be gained by using the adaptive time step.
To this end, we introduce one additional operator, \FMG{}, which is the guiding center \AMG{} operator but with a fixed time step.
The test case consists of simulating $u=5$ electrons in $\Theta = 1\times10^{-2}$ plasma until they slow down below the energy corresponding to the background temperature.
We measure how the mean slowing-down time converges when decreasing the time step in the fixed schemes or the error tolerance in the adaptive ones.
Here we noticed that the operators \FEP{} and \AMP{} converged to 0.670 s while \FMG{} and \AMG{} converged to 0.673 s.
This difference could originate from the guiding center transformation, but it is insignificant as the Coulomb logarithm is only accurate to within $1/\ln\Lambda$.
The rate of convergence for all operators is shown in Fig.~\ref{fig:benchmark}, where the error is plotted with respect to the elapsed cpu time.
In all cases, the slope of the fitted curves is approximately -1, which confirms that the operators have a weak order of convergence 1.0 as expected.
The benchmark shows that the adaptive method reduces the computational time by a factor of 10 in the particle picture, and by a factor of 3 in the guiding center picture.
The guiding center operators, both fixed and adaptive, are more efficient compared to the corresponding particle operators even though the guiding center operator has two more variables to be solved for. 

\begin{figure}[t]
\centering
\begin{overpic}[width=0.45\textwidth]{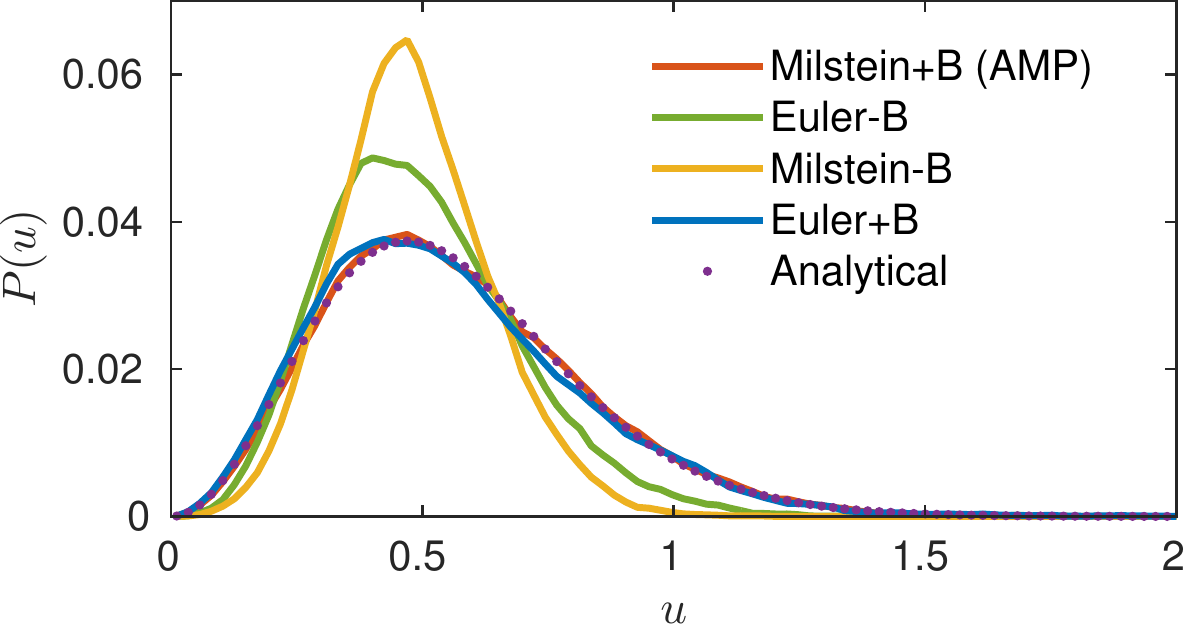}
\end{overpic}
\caption{Equilibrium momentum distribution, $P(u)$, obtained with the correct AMP operator and crippled operators compared to the analytical result.
$\pm$B indicates whether Brownian bridge was included (+) or not (-).
The simulations were done in the particle phase space.
}
\label{fig:biased dist}
\end{figure}

Now that we have verified and benchmarked the collision operators, one might wonder was it necessary to go trough all the trouble with the Milstein method and Brownian bridge?
Would the same results be achieved with using the adaptive scheme with the Euler-Maruyama method and by omitting Brownian bridge when rejecting time steps?
The answer is no as figure~\ref{fig:biased dist} clearly shows.
Using the Euler-Maruyama method and omitting Brownian bridge leads to a distribution that is strongly peaked and slightly biased to lower $u$ values.
When using the Milstein method but still omitting Brownian bridge, the bias is no longer present but the distribution remains peaked.
The peak is exactly what we would expect from earlier discussion of a biasing mechanism: the small values of $\Delta W$ are over represented as $D_\parallel$ is large, which leads to drift term being too dominant. 
This in turn drives markers towards the peak where $K$ changes sign.
When using the Euler-Maruyama method with Brownian bridge, the peak disappears but the distribution is biased, which confirms that the Euler-Maruyama method is unsuitable for adaptive time-stepping.
The difference to the analytical result is not large but it could be more significant in more complex cases than our test case.
Note that the error estimate Eq.~\eqref{eq:milstein prt ediff1} is for the Milstein method so it cannot be used when using the Euler-Maruyama method adaptively.
Instead, the ``extra'' term that separates the Euler-Maruyama and Milstein methods becomes the error estimate: $\varepsilon_\mathrm{diff} = |D_\parallel'[(\Delta W_3)^2-\Delta t]|/2\varepsilon_\mathrm{abs}$.
Therefore, using Euler-Maruyama adaptively still requires computation of $D_\parallel'$, so no computational benefits are gained when using it instead of the Milstein method.

\section{Summary and conclusion}
\label{sec:summary}

We have developed a robust and computationally efficient collision operator for simulating test particle Coulomb collisions with a background obeying Maxwell-J\"uttner statistics.
It features an adaptive time-step integration scheme and is based on the Milstein method that has both weak and strong order of convergence of 1.0.
The collision operator is fully relativistic and can operate either in the 3D particle momentum space or in the 5D guiding center phase space.
The operator was verified by showing that it converges to the known analytical results.
The adaptive scheme decreased computation time by a factor of 10 (particle momentum operator) or 3 (guiding center operator) in comparison to the fixed time step integration, when simulating slowing down of fast particles during which the collision frequency changes significantly.
The collision operator developed here is implemented in the accompanied code package which is intended to serve as a library for other codes featuring Coulomb collisions.

\section*{Acknowledgements}
\noindent
This work is part of the EUROFUSION Enabling Research project ER15-CEA-09. 
The work was partially funded by the Academy of Finland projects No. 259675 and No. 298126, and the work of E. Hirvijoki was supported by US DOE Contract DE-AC02-09-CH11466.

\appendix

\section{Langevin equation in curvilinear coordinates}
\label{app:curvilinear}

The relation between the partial differential Fokker-Planck equation and the stochastic differential Langevin equation is widely known~\cite{coffey2004applications}.
However, proofs of this relation usually consider a system in Cartesian coordinates, but the guiding center Fokker-Planck equation~\eqref{eq:gc fokker-planck} is in curvilinear coordinates.
Therefore, here we derive the relation in curvilinear coordinates to show that the guiding center Langevin equation indeed is Eq.~\eqref{eq: gc langevin}.

Let $\mathbf{z}$ be a stochastic process given by the Langevin equation
\begin{equation}
d\mathbf{z} = \mathbf{p}dt + \mathbf{g}\cdot d\W,
\end{equation}
where we assume $\W$ are independent Wiener processes.
Applying It\^o's Lemma on an arbitrary function $a(\mathbf{z})$, we get
\begin{equation}
da = \frac{\partial a}{\partial \mathbf{z}}\cdot \left(\mathbf{p}dt + \mathbf{g}\cdot d\W\right)+\frac{1}{2}\frac{\partial^2 a}{\partial \mathbf{z}\partial \mathbf{z}} : \mathbf{g}\mathbf{g}^\intercal dt.
\end{equation}
A Wiener process has the property $E[d\W] = 0$ and, therefore, taking the expectation value on both sides and formally dividing by $dt$ yields
\begin{equation}
\label{eq:appendix1}
\frac{d}{dt}E[a] = E\left[\frac{\partial a}{\partial \mathbf{z}}\cdot \mathbf{p}+\frac{1}{2}\frac{\partial^2 a}{\partial \mathbf{z}\partial \mathbf{z}} : \mathbf{g}\mathbf{g}^\intercal\right].
\end{equation}

When coordinates $\mathbf{z}$ form a curvilinear $n$-dimensional system, the differential volume element becomes $d\mathbf{z}=Jdz_1dz_2\cdots dz_n$, where $J$ is the Jacobian.
Assuming that the values of $\mathbf{z}$ obey a probability distribution $f(\mathbf{z},t)$, the left-hand side of Eq.~\eqref{eq:appendix1} becomes
\begin{align}
\frac{d}{dt}E[a] &= \frac{d}{dt}\int_\Omega af(\mathbf{z},t)Jdz_1\cdots dz_n \nonumber\\
&= \int_\Omega a\frac{df(\mathbf{z},t)}{dt}Jdz_1\cdots dz_n,
\end{align}
where the integration is over the phase space $\Omega$.
Writing the first term on the right-hand side of Eq.~\eqref{eq:appendix1} explicitly
\begin{equation}
E\left[\frac{\partial a}{\partial \mathbf{z}}\cdot \mathbf{p}\right] = \int_\Omega \frac{\partial a}{\partial \mathbf{z}}\cdot \mathbf{p}f(\mathbf{z},t)Jdz_1\cdots dz_n,
\end{equation}
and integrating by parts results in
\begin{align}
\int_\Omega \frac{\partial a}{\partial \mathbf{z}}\cdot &\mathbf{p}f(\mathbf{z},t)Jdz_1\cdots dz_n
=\int_{\partial\Omega} af(\mathbf{z},t)J\;\mathbf{p}\cdot\mathbf{\hat{n}}\;dz_1\cdots dz_n\nonumber\\
&-\int_\Omega a \frac{\partial }{\partial \mathbf{z}}\cdot\left(J\mathbf{p}f(\mathbf{z},t)\right)dz_1\cdots dz_n\nonumber\\
&=0-\int_\Omega a \frac{\partial }{\partial \mathbf{z}}\cdot\left(J\mathbf{p}f(\mathbf{z},t)\right)dz_1\cdots dz_n,
\end{align}
where the first term vanishes when we assume $f(\mathbf{z},t)\rightarrow0$ when $\mathbf{z}\rightarrow\infty$.
Now, repeating the above manipulation twice also for the second term on the right-hand side, the equation~\eqref{eq:appendix1} can be written as
\begin{align}
\int_\Omega a\frac{df(\mathbf{z},t)}{dt}&Jdz_1\cdots dz_n = \int_\Omega a\left[-\frac{\partial }{\partial \mathbf{z}}\cdot\left(J\mathbf{p}f(\mathbf{z},t)\right)\right.\nonumber\\
&\left.+\frac{1}{2}\frac{\partial}{\partial \mathbf{z}}\frac{\partial}{\partial \mathbf{z}} : \left(J\mathbf{g}\mathbf{g}^\intercal f(\mathbf{z},t)\right)\right]dz_1\cdots dz_n.
\end{align}

Since $a$ is an arbitrary function, the final step is to define $\mathbf{G}\equiv(1/2)\mathbf{g}\mathbf{g}^\intercal$ and divide by $J$ to obtain
\begin{equation}
\frac{\partial f(\mathbf{z},t)}{\partial t} = -\frac{1}{J}\frac{\partial}{\partial \mathbf{z}}\cdot\left(J\mathbf{p}f(\mathbf{z},t)\right) +\frac{1}{J}\frac{\partial}{\partial \mathbf{z}}\frac{\partial}{\partial \mathbf{z}} :\left(J\mathbf{G}f(\mathbf{z},t)\right),
\end{equation}
which is the Fokker-Planck equation in curvilinear coordinates.



\bibliographystyle{elsarticle-num}
\bibliography{bibfile}







\end{document}